\documentstyle[epsfig]{mn}

\def\gs{\mathrel{\raise0.35ex\hbox{$\scriptstyle >$}\kern-0.6em
\lower0.40ex\hbox{{$\scriptstyle \sim$}}}}
\def\ls{\mathrel{\raise0.35ex\hbox{$\scriptstyle <$}\kern-0.6em
\lower0.40ex\hbox{{$\scriptstyle \sim$}}}}

\begin{document}

\title
[Millimetre/submillimetre-wave line surveys] 
{Millimetre/submillimetre-wave emission line searches for high-redshift galaxies}  
\author
[A.\,W. Blain et al.]
{
A.\,W. Blain,$^{1, 2, 3}$ D.\,T. Frayer,$^2$ J.\,J. Bock$^4$ and N.\,Z. Scoville$^{2}$
\\ 
\vspace*{1mm}\\
$^1$ Cavendish Laboratory, Madingley Road, Cambridge, CB3 0HE.\\
$^2$ Astronomy Department, California Institute of Technology,
105-24, Pasadena, CA 91125, USA.\\
$^3$ Institute of Astronomy, Madingley Road, Cambridge, CB3 0HA.\\
$^4$ Jet Propulsion Laboratory, 4800 Oak Grove Drive, Pasadena, CA
91109, USA.\\
}
\maketitle

\begin{abstract}
The redshifted spectral line radiation emitted 
from both atomic fine-structure 
and molecular rotational transitions in the interstellar medium (ISM) of 
high-redshift galaxies can be detected in the centimetre, 
millimetre and submillimetre 
wavebands. Here we predict the counts of galaxies detectable in an 
array of molecular and atomic lines. This calculation requires a reasonable 
knowledge of both the surface density of these galaxies on the sky, 
and the physical 
conditions in their ISM. The surface density is constrained 
using the results of submillimetre-wave continuum surveys. 
Follow-up OVRO Millimeter Array observations of two of the galaxies 
detected in the dust 
continuum have provided direct 
measurements of CO rotational line emission at redshifts of 2.56 and 2.81. 
Based on these direct high-redshift observations and on models of the 
ISM that are 
constrained by observations of low-redshift ultraluminous infrared 
galaxies, we 
predict the surface density of line-emitting galaxies as a function of 
line flux 
density and observing frequency. 
We incorporate the sensitivities and
mapping speeds of existing and future millimetre/submillimetre-wave
telescopes and spectrographs, and so
assess the prospects 
for blank-field surveys to detect this line emission 
from gas-rich high-redshift galaxies. 
\end{abstract}  


\section{Introduction}

The redshifted far-infrared/submillimetre-wave line emission from the 
interstellar medium (ISM) in galaxies could be exploited to detect new 
samples of distant gas-rich galaxies and active galactic nuclei (AGN) 
(Loeb 1993; Blain 1996; van der Werf \& Israel 1996; Silk \& Spaans 1997; 
Stark 1997; Combes, Maoli \& Omont 1999; van der Werf 1999). This emission is 
attributable both to molecular rotational transitions, in particular from 
carbon monoxide (CO), and to atomic fine-structure transitions, in particular 
from the singly ionized 158-$\mu$m carbon [C{\sc ii}] line.

Redshifted CO emission has been detected from a range of known 
high-redshift galaxies and quasars in the millimetre/submillimetre waveband, as 
summarized by Frayer et al.\ (1998) and Combes et al.\ (1999). Many of these 
galaxies are known to be gravitationally lensed by a foreground galaxy, an effect 
which potentially complicates the interpretation of the results by altering the 
ratios of the inferred luminosities in the continuum and the detected lines 
(Eisenhardt et al.\ 1996). In only a small subsample of the detected
galaxies (Solomon, Downes \& Radford 1992; Barvainis et al.\ 1997; Downes 
et al.\ 1999) have multiple lines been detected, providing an 
opportunity to investigate the astrophysics of the ISM. 

So far there have been very few detections of redshifted fine-structure lines, 
despite careful searches, for example, for both [C{\sc ii}] (Isaak et al.\ 1994; Ivison, 
Harrison \& Coulson 1998a; van der Werf 1999) and singly ionized 
205-$\mu$m [N{\sc ii}] emission (Ivison \& Harrison 1996). Neutral carbon 
[C{\sc i}] emission, which is considerably less intense than [C{\sc ii}] and 
[N{\sc ii}] emission in the Milky Way (Wright et al.\ 1991) and nearby galaxies 
(Stacey et al.\ 1991), has been detected from the gravitationally lensed 
Cloverleaf quasar (Barvainis et al.\ 1994). The most luminous high-redshift 
galaxies and quasars have necessarily been targetted in these searches. 

[C{\sc ii}] fine-structure emission is powerful in both the Milky Way and in 
sub-$L^*$ galaxies, in which it accounts for about 0.5\,per cent of the 
bolometric far-infrared luminosity (Nikola et al.\ 1998). 
However, based on observations of a 
limited number of low-redshift galaxies using the {\it Infrared Space 
Observatory (ISO)} (Malhotra et al.\ 1997; Luhman et al.\ 1998; Pierini et al.\ 
1999), it appears that a systematically lesser fraction of the bolometric 
luminosity of more luminous galaxies appears as [C{\sc ii}] emission, 
about 0.1\,per cent (Luhman et al.\ 1998). As 
noted by Luhman et al.\ (1998) and van der Werf (1999), the results of 
these {\it ISO} observations are fully consistent with the non-detection of 
redshifted 
fine-structure emission from high-redshift galaxies using ground-based 
submillimetre-wave telescopes. The results of Kuiper Airborne Observatory 
(KAO) observations of the Galactic centre (Erickson et al.\ 1991) indicate that 
63- and 146-$\mu$m neutral oxygen [O{\sc i}] fine-structure emission becomes 
steadily more luminous as compared with that from [C{\sc ii}] as the far-infrared 
luminosity of gas clouds increases. However, currently there is insufficient 
published data available to address this issue in external galaxies. 

We attempt here to predict the counts of distant gas-rich line-emitting 
galaxies that could be detected in the millimetre/submillimetre waveband. 
There are two challenges to making reliable predictions. First, there are 
limited data available from which to construct a clear understanding of the  
astrophysics of the ISM in high-redshift galaxies. There are only a few tens of 
detections of line emission from these objects, the majority of which have 
been made in galaxies that are gravitationally lensed by foreground galaxies. 
Because of the potential for differential magnification across and within the 
lensed galaxy, neither the ratios of the line and continuum 
luminosities nor the excitation conditions 
in the ISM are known accurately in these cases. As shown by {\it ISO} 
[C{\sc ii}] observations, extrapolation of the observed properties of low-redshift 
galaxies with relatively low luminosities to greater luminosities in high-redshift 
galaxies is not necessarily reliable. Secondly, the space density and form of 
evolution of gas-rich galaxies at high redshifts has not been well determined.
Thus the existing predictions of the observability of high-redshift 
submillimetre-wave line emission have concentrated on either discussing the 
potential observability of individual high-redshift galaxies (van der Werf \& 
Israel 1996; Silk \& Spaans 1997; Combes et al.\ 1999; van der Werf 1999), or 
have relied on extensive extrapolations, from the populations of low-redshift 
ultraluminous infrared galaxies (ULIRGs) (Blain 1996) and from 
low-redshift $L^*$ bulges to the properties of proto-quasars at $z \sim 10$ 
(Loeb 1993). 

Both of these difficulties can be addressed by exploiting the results of 
deep 450- and 850-$\mu$m dust continuum radiation surveys made using the 
Submillimetre Common-User Bolometer Array (SCUBA) camera 
(Holland et al.\ 1999) at the James Clerk Maxwell Telescope (JCMT; 
Smail, Ivison \& 
Blain 1997; Barger et al.\ 1998; Hughes et al.\ 1998; Barger, Cowie \& 
Sanders 1999; Blain et al.\ 1999b, 
2000; 
Eales et al.\ 1999). These surveys are sensitive to galaxies 
at very high redshifts (Blain \& Longair 1993), and have detected a considerable 
population of very luminous dust-enshrouded galaxies. The 15-arcsec 
angular resolution of the JCMT is rather coarse, but 
reliable identifications can be made              
by combining the SCUBA 
images with multi-waveband follow-up images and spectra 
(Ivison et al.\ 1998b, 2000; 
Smail et al.\ 
1998, 2000; Barger et al.\ 1999b; Lilly et al. 1999), 
and crucially with observations of 
redshifted CO emission, which are 
currently available for two 
submillimetre-selected galaxies:   
SMM\,J02399$-$0136 at $z=2.81$  
and SMM\,J14011+0252 at $z=2.56$ (Frayer et al.\ 1998, 1999; 
Ivison et al.\ 1998b, 2000). 
The bolometric and CO-line luminosities of these 
galaxies are reasonably well known, and because they are lensed by clusters 
rather than individual foreground galaxies, their 
inferred line ratios are not subject to modification
by lensing. These observations 
thus provide a useful template with which to describe the properties of the 
ISM and line emission in high-redshift, dust-enshrouded, gas-rich galaxies.  

In Section 2 we discuss the existing line observations, and summarize our 
current state of knowledge about the evolution and redshift distribution of 
galaxies that have been discovered in submillimetre-wave dust continuum 
surveys. In Section\,3 we describe our model of line emission from these 
galaxies, and present the results, as based on our understanding of 
high-redshift continuum sources. In Section\,4 we discuss the observability of 
this hypothetical population using existing and future 
millimetre/submillimetre-wave spectrographs. Unless otherwise stated, we
assume that $H_0 = 50$\,km\,s$^{-1}$\,Mpc$^{-1}$, $\Omega_0=1$ and 
$\Omega_\Lambda =0$. 

\section{Background information} 

\subsection{Line observations} 

Ground-based 
telescopes have detected 
molecular rotation lines and atomic fine-structure lines 
from low-redshift galaxies (e.g., 
Sanders et al.\ 1986; Wild et al.\ 1992; 
Devereux et al.\ 1994; Gerin \& Phillips 1999; Mauersberger et al.\ 1999). 
Atomic fine-structure lines have also been observed from bright galactic 
star-forming regions and nearby galaxies using the KAO (Stacey et al.\ 1991; 
Nikola et al.\ 1998), {\it COBE} (Wright et al.\ 1991) and {\it ISO} 
(Malhotra et al.\ 1997; Luhman et al.\ 1998; Pierini et al.\ 1999). 

CO rotational line emission has been detected 
successfully from various 
high-redshift 
galaxies and quasars, including 
the first identified high-redshift 
ULIRG {\it IRAS} F10214+4724 (Solomon et al.\ 1992), the 
gravitationally lensed Cloverleaf quasar H\,1413+117 
(Barvainis et al.\ 1994; Kneib et al.\ 
1998), various quasars at $z \simeq 4$, including BR\,1202$-$0725 
(Ohta et al.\ 1996, 1998; Omont et al.\ 1996) and the extremely luminous 
APM\,08279+5255 (Lewis et al.\ 1998; Downes et al.\ 1999), and the 
submillimetre-selected galaxies SMM\,J02399$-$0136 and SMM\,J14011+0252 
(Frayer et al.\ 1998, 1999). 
A significant fraction of the dynamical mass in 
many of these systems is inferred to be in the form of molecular gas, and it 
is plausible that they are observed in the process of forming the bulk of 
their stellar populations.

\subsection{Atomic fine-structure lines} 

Atomic fine-structure lines emitted at wavelengths longer than about 
100\,$\mu$m -- [C{\sc ii}] at 1900\,GHz, [N{\sc ii}] at 1460 and 2460\,GHz, 
[O{\sc i}] at 2060\,GHz, and [C{\sc i}] at 492 and 809\,GHz -- are redshifted 
into atmospheric windows for galaxies at redshifts $z \ls 5$, 
the redshift range within which at least
80\,per cent of dust-enshrouded galaxies detected by SCUBA appear to lie
(Smail et al.\ 1998; Barger et al.\ 1999b; Lilly et al.\ 1999).
There are many other 
mid-infrared lines with shorter restframe emission wavelengths 
(see, e.g., Lutz et al.\ 1998); however, unless massive galaxies exist at 
$z \sim 10$, these lines will not be redshifted into atmospheric
windows accessible to ground-based telescopes. 

Here we assume that the line-to-bolometric luminosity ratio 
$f_{\rm line} = 10^{-4}$ for the [C{\sc ii}] line in all high-redshift 
gas-rich galaxies,
corresponding to the value observed in low-redshift ULIRGs. 
The equivalent value for [C{\sc i}]$_{\rm 492\,GHz}$, 
$f_{\rm line} = 2.9 \times 10^{-6}$ is chosen to match the value observed 
by Gerin \& Phillips (1999) in Arp\,220. 
The values of $f_{\rm line}$ for 
other fine-structure lines listed in Table\,1 are chosen by scaling the results 
of observations of the Milky Way and low-redshift galaxies (Genzel et al.\ 1990; 
Erickson et al.\ 1991; Stacey et al.\ 1991; Wright et al.\ 1991). This
approach should lead to a reliable estimate of the counts of [C{\sc i}] 
and [C{\sc ii}] lines, 
but greater uncertainty in the N and O line predictions. In the 
absence of more observational data, which would ideally allow luminosity 
functions to be derived for each line, we stress that the predictions of the 
observability of redshifted fine-structure lines made in Section\,3 must 
be regarded as tentative and preliminary. 

\subsection{CO rotational transitions} 

\subsubsection{CO line excitation}
 
Much more observational data is available about the properties of the ladder
of CO rotational transitions in the ISM. The energy of the $J$th level in the 
CO 
molecular rotation ladder $E_J = k_{\rm B} T$, where $T = J (J+1)$[2.77\,K], 
and so the energy of a photon produced in the $J+1 \rightarrow J$ transition 
is $h \nu = k_{\rm B} (J+1)$[5.54\,K]. The population of the $J$ states can be 
calculated by assuming a temperature and density for the emitting gas. The 
primary source of excitation is expected to be collisions with molecular 
hydrogen (H$_2$), which 
dominates the mass of the ISM, with a role for radiative excitation, 
including that attributable to the cosmic microwave
background radiation (CMBR). 
By taking into account the spontaneous 
emission rate, $A_{J+1, J} \propto {\nu^3} (J+1) /(2J+3)$ 
with $A_{1,0} = 6 \times 10^{-8}$\,s$^{-1}$, and details of the 
optical depth and geometry of gas and dust in the emitting region, the 
luminosities of the various $J+1 \rightarrow J$ rotational transitions can be 
calculated.  
If the $J$ state is to be thermally
populated, then the rate of CO--H$_2$ collisions in the ISM gas must be greater 
than about $A_{J+1, J}^{-1}$. This condition will not generally be met for a 
temperature of 50\,K in the CO(5$\rightarrow$4) transition unless the  
density of H$_2$ molecules exceeds about  
$2 \times 10^5$\,cm$^{-3}$, which is many times denser than the 
$10^4$\,cm$^{-3}$ that appears to be 
typical of low-redshift ULIRGs (Downes \& Solomon 1998). 
Radiative excitation 
and optical depth effects, perhaps in very non-isotropic geometries, with very 
pronounced substructure, will complicate the situation greatly in real galaxies.
In general, calculations 
of level populations are very complex, 
and at high redshifts there are very few 
data with which to constrain models. 

Probably the best way to investigate the 
conditions in very luminous distant galaxies is to study their low-redshift 
ULIRG counterparts, and the rare 
examples of high-redshift galaxies and quasars for which more than one CO 
transition has been detected (e.g. Downes et al.\ 1999), bearing in 
mind the potential effects of gravitational lensing. 

In order to try and make reasonable predictions for the line ratios in 
high-redshift galaxies, we employed a standard large velocity 
gradient (LVG) analysis (e.g. de Jong, Dalgarno \& Chu 1975) to 
estimate how the CO line ratios are affected by the temperature, density and 
finite spatial extent of the ISM, and by the radiative excitation caused by the 
CMBR. In the third column of Table\,1 we show the results 
from this model, assuming a density of 10$^4$\,cm$^{-3}$, 
which is typical of the central regions of ULIRGs 
(Downes \& Solomon 1998), and a standard value of 
$X({\rm CO})/({\rm d}v/{\rm d}r) = 3 \times 10^{-5}\,({\rm km}\,{\rm s}^{-1}\,{\rm 
pc}^{-1})^{-1}$. We assume a kinetic temperature of 
53\,K, which is the temperature of the dominant cool dust component in 
the SCUBA galaxy SMM\,J02399$-$0136 (Ivison et al.\ 1998b).
Higher dust 
temperatures of about 80 and 110\,K are inferred for other well-studied 
high-redshift galaxies {\it IRAS} F10214+4724 and APM\,08279+5255 respectively, 
but these are very exotic galaxies, and the results are potentially 
modified by 
the effects of differential gravitational lensing. 
We assume a background 
temperature of 10\,K, the temperature of the CMBR at $z=2.7$, 
the mean redshift of the two SCUBA galaxies with CO detections. 
The line-to-continuum bolometric luminosity 
ratio $f_{\rm line} = L_{J+1 \rightarrow J} / 
L_{\rm FIR}$ can be calculated if the bolometric continuum luminosity 
$L_{\rm FIR}$ is known. 
There is a clear trend of a reduction in the CO 
line-to-bolometric luminosity ratio of luminous 
infrared galaxies as the bolometric luminosity increases, with a large scatter, 
which is consistent with $f_{\rm line} \propto L_{\rm FIR}^{-0.5}$
(Sanders et al.\ 1986). We normalize 
the results to the observed CO(3$\rightarrow$2) line 
luminosity in the $L_{\rm FIR} \simeq
10^{13}$\,L$_\odot$ SCUBA galaxies 
SMM\,J02399$-$0136 and SMM\,J14011+0252, in which the ratios 
of the luminosity in the CO(3$\rightarrow$2) line to $L_{\rm FIR}$ is about 
$2.1 \times 10^{-6}$ and $5.3 \times 10^{-6}$ respectively (Frayer et al.\ 1998,
1999), with errors of 
order 50\,per cent. The dominant source of error is the uncertainty in the 
bolometric luminosity. 

We compare the results obtained in a simple 
equilibrium case, in which the density of CO molecules, the 
spontaneous emission rates and the transition energies in different $J$ 
states are multiplied to give  
the luminosity in each transition, 
\begin{equation}
L_{J+1 \rightarrow J} \propto \nu^3 (J+1)^2 
\exp{ \{ - [2.77\,{\rm K}] (J+1)(J+2)/T} \}. 
\end{equation}
The results are shown in 
the fourth and fifth columns of Table\,1 
for $J \le 9$, assuming kinetic temperatures of 
38 and 53\,K respectively, the temperature 
generated by simple fits 
to the observed counts of {\it IRAS} and {\it ISO} galaxies 
(Blain et al.\ 1999c), and the spectral energy distribution (SED) of  
SMM\,J02399$-$0136. 

\begin{table*}
\caption{The fraction of the bolometric luminosity of distant dusty galaxies 
that is assumed to be emitted in a variety of submillimetre-wave lines 
$f_{\rm line}$, their restframe emission frequencies $\nu_{\rm rest}$, and the 
redshifts 
at which the lines would be detected in the important observing bands at 90, 
230, 345 and 650\,GHz; $z_{90}$, $z_{230}$, $z_{345}$ and $z_{650}$ respectively. 
The CO ratios $f_{\rm line}$ listed in the third, fourth and fifth 
columns are calculated assuming the LVG model described in Section\,2, and in 
two local thermal equilibrium models with kinetic 
temperatures of 38 and 53\,K respectively. 
The fraction of the bolometric luminosity of the galaxies in all CO transitions,
obtained by adding all values of $f_{\rm line}$ is also shown. 
The values of $f_{\rm line}$ listed for fine-structure 
transitions in the final six rows are derived from observations of low-redshift 
galaxies and the Milky Way: see Section\,2.2. The line styles and thicknesses 
used to represent the various transitions are listed in the final column.}
{\vskip 0.75mm}
{$$\vbox{
\halign {\hfil #\hfil && \quad \hfil #\hfil \cr
\noalign{\hrule \medskip}
Line & $\nu_{\rm rest}$ & $f_{\rm line}$ & $f_{\rm line}$ & $f_{\rm line}$ & 
$z_{90}$ & $z_{230}$ & $z_{345}$ & $z_{650}$ & Line style\cr
 & / GHz & (LVG) & ($T=38$\,K) & ($T=53$\,K) & & & & \cr 
\noalign{\smallskip \hrule \medskip}
CO(1$\rightarrow$0) & 115 & 
$1.8 \times 10^{-7}$ & $3.4 \times 10^{-8}$ & $2.8 \times 10^{-8}$ & 
0.28 & N/A & N/A & N/A & CO lines are represented\cr
CO($2\rightarrow 1$) & 230 & 
$1.3 \times 10^{-6}$ & $8.2 \times 10^{-7}$ & $7.2 \times 10^{-7}$ & 
1.6 & 0.0 & N/A & N/A & by solid lines whose \cr
CO($3\rightarrow 2$) & 345 & 
$4.0 \times 10^{-6}$ & $4.0 \times 10^{-6}$ & $4.0 \times 10^{-6}$ & 
2.8 & 0.50 & 0.0 & N/A & thickness increases with \cr
CO($4\rightarrow 3$) & 461 & 
$8.3 \times 10^{-6}$ & $9.6 \times 10^{-6}$ & $1.1 \times 10^{-5}$ & 
4.1 & 1.0 & 0.34 & N/A & the value of $J$ \cr
CO($5\rightarrow 4$) & 576 & 
$1.3 \times 10^{-5}$ & $1.4 \times 10^{-5}$ & $2.0 \times 10^{-5}$ & 
5.4 & 1.5 & 0.67 & N/A & \cr
CO($6\rightarrow 5$) & 691 & 
$1.6 \times 10^{-5}$ & $1.4 \times 10^{-5}$ & $2.6 \times 10^{-5}$ & 
6.7 & 2.0 & 1.0 & 0.06 & \cr
CO($7\rightarrow 6$) & 806 & 
$1.4 \times 10^{-5}$ & $1.1 \times 10^{-5}$ & $2.8 \times 10^{-5}$ & 
8.0 & 2.5 & 1.4 & 0.24 & \cr
CO($8\rightarrow 7$) & 922 & 
$4.2 \times 10^{-6}$ & $6.8 \times 10^{-6}$ & $2.3 \times 10^{-5}$ & 
9.2 & 3.0 & 1.7 & 0.42 & \cr
CO($9\rightarrow 8$) & 1040 & 
$1.6 \times 10^{-9}$ & $3.3 \times 10^{-6}$ & $1.6 \times 10^{-5}$ & 
12 & 3.5 & 2.0 & 0.60 & \cr
\noalign{\smallskip} 
Total CO & N/A & $6.1 \times 10^{-5}$ & $6.6 \times 10^{-5}$ & 
$1.5 \times 10^{-4}$ & N/A & N/A & N/A & N/A & N/A \cr
\noalign{\medskip} 
[C\sc{ii}] & 1890 & $10^{-4}$ & ... & ... & 20 & 7.2 & 4.5 & 1.9 & 
Thick dashed line\cr
[C\sc{i}]$_{809\,{\rm GHz}}$ & 809 & $2.9 \times 10^{-5}$ & ... & ... &   
8.0 & 2.5 & 1.3 & 0.24 & Thick dot-dashed line\cr
[C\sc{i}]$_{492\,{\rm GHz}}$ & 492 & $2.9 \times 10^{-6}$ & ... & ... &  
4.5 & 1.1 & 0.43 & N/A & Thin dot-dashed line\cr
\noalign{\medskip} 
[N\sc{ii}]$_{2460\,{\rm GHz}}$ & 2460 & $6.4 \times 10^{-4}$ & ... & ... &  
26 & 9.6 & 6.1 & 2.8 & Thick dotted line\cr
[N\sc{ii}]$_{1460\,{\rm GHz}}$ & 1460 & $4.0 \times 10^{-4}$ & ... & ... &  
15 & 5.3 & 4.2 & 1.2 & Thin dotted line\cr
[O\sc{i}]$_{2060\,{\rm GHz}}$ & 2060 & $2.4 \times 10^{-4}$ & ... & ... &  
22 & 8.0 & 5.0 & 2.2 & Thin dashed line\cr 
\noalign{\smallskip \hrule}
\noalign{\smallskip}\cr}}$$}
\end{table*}

\subsubsection{The effects of different excitation conditions} 

The values of $f_{\rm line}$ listed in Table\,1 for CO transitions in the 
LVG and the 38- and
53-K thermal 
equilibrium models differ. However, in lines with 
$J \ls 7$ the differences are less than a factor of a few. 
Given the current 
level of uncertainty in the data that support these calculations, this is an 
acceptable level. The differences between the results are more marked 
at large values of $J$. The consequences of these differences 
for the key 
predictions of the source counts of line-emitting galaxies are 
discussed in Section\,3.2.

Throughout the paper we use the LVG model to describe the CO line 
emission of dusty galaxies. Observations of
the CO line ratios in low-redshift dusty galaxies indicate a wide 
range of excitation conditions, 
$T_{\rm b}$[CO(3$\rightarrow$2)/CO(1$\rightarrow$0)] $\sim0.2$--1
(Mauersberger et al.\ 1999). In the central regions of starburst 
nuclei this temperature ratio tends to be systematically higher, with 
$T_{\rm b}$[CO(3$\rightarrow$2)/CO(1$\rightarrow$0)] $\simeq 0.5$--1 
(Devereux et al.\ 1994). In our chosen LVG model, listed in 
Table\,1, this ratio is $\simeq 0.9$, which is consistent with the 
observations of the central regions of M82 (Wild et al.\ 1992) and 
Arp 220 (Mauersberger et al.\ 1999). The value of the $X$ parameter 
in the model has little effect on these ratios; however, reducing the 
density from 10$^4$ to 3300 and 1000\,cm$^{-3}$ reduces the 
predicted ratio to 0.81 and 
0.51 respectively. Our model looks reasonable in the light of these 
observations, as the high-redshift galaxies would typically be expected 
to be ULIRGs with high gas densities. 

There have been two recent discussions of the observability of CO line 
emission from high-redshift galaxies. Silk \& Spaans (1997) describe the 
effect of the increasing radiative excitation of high-$J$ lines at very 
high redshifts 
because of the increasing temperature of the CMBR. 
The median 
redshift of the SCUBA galaxies is likely to be about 2--3 (Barger et al.\ 
1999b; Smail et al.\ 2000; Lilly et al.\ 1999), with perhaps 10--20\,per cent at 
$z \gs 5$, and because  
there is currently no strong evidence for the existence of a 
large population of metal-rich galaxies at $z\gs10$, this effect 
is unlikely to be very important. Combes et al.\ (1999) include a hot dense 
90-K/10$^6$\,cm$^{-3}$ component in the ISM of their 
model high-redshift galaxies, in addition to the cooler less dense component 
included in our models, and conduct LVG calculations to determine CO 
emission-line
luminosities. Understandably, the luminosity of high-$J$ CO lines is 
predicted to be greater in their models as compared with the values listed in 
Table\,1. The 
continuum SED of the best studied SCUBA galaxy SMM\,J02399$-$0136 
certainly includes a contribution from dust at temperatures greater than 53\,K, 
but here we avoid including additional hot dense phases of the ISM in our 
models in order to avoid complicating the models and to 
try and make conservative predictions for the observability 
of high-$J$ CO lines at high redshift. 

Only additional observations of CO in high-redshift galaxies will allow 
us to improve the accuracy of the conditions in the ISM that are assumed in 
these models. The detection of the relative intensities of the 
CO(9$\rightarrow$8) and CO(5$\rightarrow$4) emission 
from APM\,08279+5255 (Downes et al.\ 1999), and the  
ratio of the intensities of the multiple CO lines detected in 
BR\,1202$-$0725 at $z=4.7$ (Ohta et al.\ 1996, 1998; Omont et al.\ 1996) are 
broadly consistent with the values of
$f_{\rm line}$ listed in the 53-K thermal equilibrium model: see Table\,1.

The evolution of the abundance of CO and dust throughout an episode of 
star formation activity have been investigated by Frayer \& Brown (1997), and 
the detailed appearance of the submillimetre-wave emission-line 
spectrum of galaxies requires 
a careful treatment of the radiative transfer between stars, AGN, gas and dust
in an appropriate geometry. However, given that the amount of information on the 
spectra of high-redshift galaxies is currently not very great, it seems sensible to 
base estimates of the properties of line emission on the template of the
submillimetre-selected galaxies studied by Frayer et al.\ (1998, 1999).   

\subsubsection{Other molecular emission lines} 

There could also be a contribution from rotational lines emitted by other 
species, such as NH$_3$, CS, HCN, HCO$^+$ and H$_2$O; however, it 
seems unlikely that these emission lines would dominate the energy 
emitted in CO unless the densities and excitation temperatures are very high.

\subsection{High-redshift dusty galaxies} 

The surface density of 850-$\mu$m SCUBA galaxies is now known reasonably 
accurately between flux densities of 1 and 10\,mJy (Barger et al.\ 1999a; 
Blain et al.\ 1999b, 
2000). By 
combining knowledge of the properties of the SCUBA galaxies, the low-redshift 
60-$\mu$m {\it IRAS} galaxies (Saunders et al.\ 1990; Soifer \& Neugebauer 
1991), the 175-$\mu$m {\it ISO} galaxies 
(Kawara et al.\ 1998; Puget et al.\ 1999) and the intensity of  
far-infrared background radiation 
(Fixsen et al.\ 1998; Hauser et al.\ 1998; 
Schlegel, Finkbeiner \& Davis 1998), it is 
possible to construct a series of self-consistent models that can account for 
all these data (Guiderdoni et al.\ 1998; Blain et al.\ 1999b,c), under various 
assumptions about the formation and evolution of galaxies. The `Gaussian' 
model, which is described by Blain et al.\ (1999b) and based on pure luminosity 
evolution of the low-redshift 60-$\mu$m luminosity function (Saunders et al.\ 
1990), was modified slightly to take account of the tentative redshift 
distribution derived from the SCUBA lens survey by Barger et al.\ (1999b). 
This 
`modified Gaussian' model is used as a base for the predictions of line 
observability presented here. The evolution function increases as 
$(1+z)^\gamma$ with $\gamma \simeq 4$ at $z \ls 1$, has 
a 1.0-Gyr-long Gaussian burst of luminous activity centred at $z=1.7$, 
in which $L^*$ is 70 times greater than the value of $L^*$ at
$z=0$ (Saunders et al.\ 1990), and then declines at $z \gs 3$.

This model contains fewer parameters than there are 
constraining pieces of information, and so should provide a reasonable 
description of the properties of high-redshift dust-enshrouded galaxies, 
which are expected to be the most easily detectable sources of line 
emission. Future observations will inevitably provide more 
information and demand modifications to the model of galaxy evolution
discussed above; however, it is likely to provide a sound basis for the 
predictions below.

\section{Line predictions}

In this paper we are concerned with the detectability of redshifted lines rather 
than with their resolution. As a result, we want to estimate the total luminosity 
of a line, and will not be concerned with details of its profile. The results are 
all presented as integrated flux densities, determined over the whole line 
profile. Where relevant, a line width of 300\,km\,s$^{-1}$ is 
assumed. 

\begin{figure*}
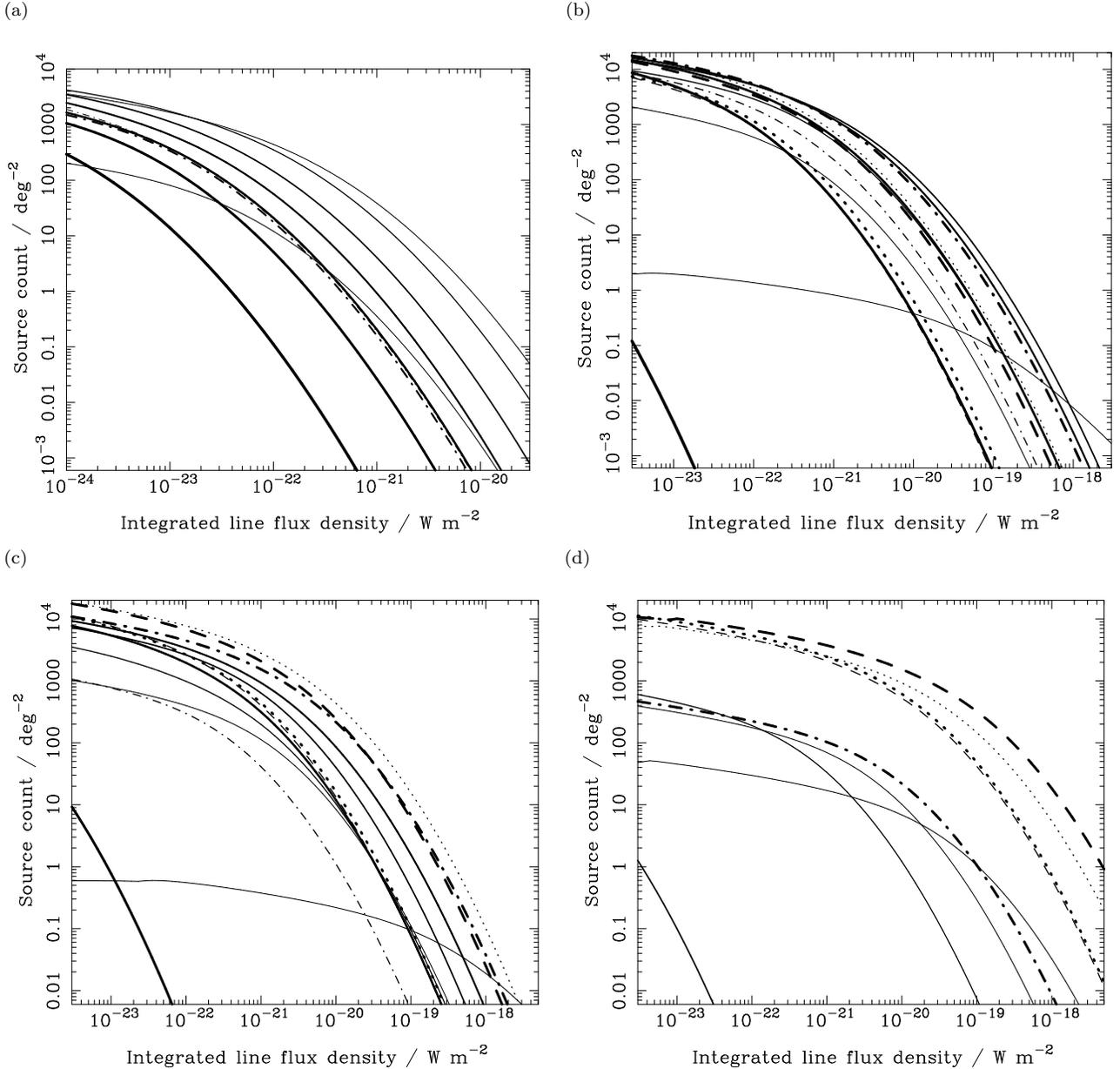

\begin{minipage}{170mm}
(a) \hskip 81mm (b)
\begin{center}
\epsfig{file=rev90.ps, width=7.35cm, angle=-90} \hskip 5mm
\epsfig{file=rev230.ps, width=7.35cm, angle=-90}
\end{center}
(c) \hskip 81mm (d)
\begin{center}
\epsfig{file=rev345.ps, width=7.15cm, angle=-90} \hskip 5mm
\epsfig{file=rev650.ps, width=7.15cm, angle=-90}
\end{center}
\caption{Predicted counts of CO rotation and fine-structure lines detectable 
from galaxies 
in a 
1-GHz band centred at 90\,GHz (a), and in 8-GHz bands centred on (b) 230\,GHz, 
(c) 345\,GHz and (d) 650\,GHz. The band in (a) is 
accessible to the existing 
BIMA, IRAM, Nobeyama and OVRO interferometer arrays. These 
arrays also operate with a narrower 1-GHz bandwidth around 
230-GHz (b). ALMA will operate in all the bands shown in (b), (c) and 
(d), with a current goal bandwidth of 16\,GHz (Wootten 2000). SPIFI and the 
HIFI and SPIRE-FTS instruments 
on the {\it FIRST} satellite will operate in the 
band shown in (d). 
The line styles and thicknesses 
correspond to each transition listed in Table\,1. 
In (a) the CO rotational transitions from 
CO(1$\rightarrow$0) to
CO(8$\rightarrow$7) and the [C{\sc i}]$_{\rm 492\,GHz}$ line are 
present, in 
(b)
CO(2$\rightarrow$1) to
CO(9$\rightarrow$8) and all the fine-structure lines listed in Table\,1 
are detectable, in (c) CO(3$\rightarrow$2) to CO($9\rightarrow$8)
and all the fine-structure lines are present, and in (d) 
the CO
transitions from CO(6$\rightarrow$5) to CO(9$\rightarrow$8) and all but the 
[C{\sc i}]$_{\rm 492\,GHz}$ fine-structure line 
are present. 
$10^{-20}$\,W\,m$^{-2}$ is equivalent to 3.3, 1.3, 0.9 and 
0.5\,Jy\,km\,s$^{-1}$ 
at 90, 230, 345 and 650\,GHz respectively.
}
\end{minipage}
\end{figure*} 

The evolution of the 60-$\mu$m luminosity function of 
dusty galaxies (Saunders et al.\ 1990) 
is assumed to be defined 
by the modified Gaussian model. 
The bolometric far-infrared continuum luminosity function 
$\Phi_{\rm bol}(L_{\rm FIR})$ can then be calculated by integrating over a 
template dusty galaxy SED (Blain et al.\ 1999b). $\Phi_{\rm bol}$ can 
in turn be 
converted into a luminosity function for each line 
listed in Table\,1, 
$\Phi_{\rm line}(L_{\rm line}, z)$, by evaluating $\Phi_{\rm bol}$ at 
the bolometric luminosity $L_{\rm FIR}$ that corresponds to the line 
luminosity $L_{\rm line}$. 

Based on observations by Sanders et al.\ 
(1986), the ratio of the luminosity in the CO(1$\rightarrow$0) line, 
$L_{\rm CO}$, to 
$L_{\rm FIR}$ is a function of $L_{\rm FIR}$, with 
$L_{\rm CO} \propto L_{\rm FIR}^{0.5}$. Hence, when making the 
transformation 
from $L_{\rm line}$ to $L_{\rm FIR}$ for the CO lines listed in 
Table\,1, we use the relationship 
$L_{\rm FIR} = L_{\rm line} / f_{\rm line}$, in which $f_{\rm line} 
\propto L_{\rm FIR}^{-0.5}$ and is normalized to the  
value listed in Table\,1 
at $10^{13}$\,L$_\odot$, the 
luminosity of the submillimetre-selected 
galaxies SMM\,J02399$-$0136 and SMM\,J14011+0252. 
There is no evidence from 
{\it ISO} observations of any systematic luminosity 
dependence in the value of $f_{\rm line}$ for [C{\sc ii}] emission from 
ULIRGs, 
and so we assume that the 
values of $f_{\rm line}$ listed in Table\,1 describe the line-to-bolometric 
luminosity relation in the fine-structure lines at all luminosities, 
that is $L_{\rm FIR} = L_{\rm line} / f_{\rm line}$, where 
$f_{\rm line}$ is constant. 

The surface density of line-emitting galaxies $N(>S)$ that can be detected at 
integrated flux densities brighter than $S$, as measured in Jy\,km\,s$^{-1}$ or 
W\,m$^{-2}$, in an observing band spanning the frequency range between 
$\nu_{\rm obs}$ and $\nu_{\rm obs} + \Delta \nu_{\rm obs}$  
can be calculated by integrating the luminosity function of a line 
$\Phi_{\rm line}$ over the 
redshifts for which the line is in the 
observing band, and luminosities $L_{\rm line}$ 
greater than the detection limit 
$L_{\rm min}(S, z)$. The count of 
galaxies is thus  
\begin{equation} 
N(>S) = \int_{z_1}^{z_2} \int_{L_{\rm min}(S, z)}^\infty 
\!\!\!\!\!\!\!\!
\Phi_{\rm line}(L_{\rm line}, z) \, {\rm d}L_{\rm line} \, 
D^2(z) { { {\rm d}r } \over { {\rm d}z } } \, {\rm d}z.  
\end{equation}
$D$ is the comoving distance parameter to redshift $z$, and ${\rm d}r$ is the 
comoving distance element. For a line with a restframe emission frequency 
$\nu_{\rm line}$, the limits in equation (2) are
\begin{equation} 
z_1 = { {\nu_{\rm obs} + \Delta \nu_{\rm obs}} \over {\nu_{\rm line}} } - 1, 
\end{equation} 
and  
\begin{equation}
z_2 = { {\nu_{\rm obs}} \over {\nu_{\rm line}} } -1. 
\end{equation}
If $z_1 \ge z_0$, where $z_0$ is the maximum redshift assumed for the 
galaxy population, or if $z_2 \le 0$, then the count $N=0$. Here $z_0=10$ is 
assumed. 
The minimum detectable line luminosity, 
\begin{equation} 
L_{\rm min}(S, z) = 4 \pi S D^2 (1+z)^2. 
\end{equation}
If the integrated flux density of a line $S$ observed at frequency $\nu$, is 
1\,W\,m$^{-2}$, then the same quantity can be expressed as 
$3(\nu/{\rm Hz})^{-1} \times 10^{31}$\,Jy\,km\,s$^{-1}$.

The counts calculated for the transitions listed in Table\,1 are shown in 
Figs\,1--3. Predictions are made at centre frequencies (bandwidths) of 
23 (8) -- the radio K band, 50 (8) -- the radio Q band, 90 
(1), 230 (8), 345 (8), 650 (8), 200 (100)\,GHz, and 1 (1)\,THz, 
in Figs 2(a), 2(b), 
1(a), 1(b), 1(c), 1(d), 3(a) and 3(b) 
respectively. The 8-GHz radio bandwidth is matched 
to the performance specified for the upgraded VLA.
The 1-GHz bandwidth at 90\,GHz is approximately matched to 
the current performance of millimetre-wave interferometer arrays. The 
8-GHz bandwidth 
at 230, 345 
and 650\,GHz matches that for the SPIFI Fabry--Perot 
spectrograph (Stacey et al.\ 1996; Bradford et al.\ 1999, in preparation) 
and is a plausible value for the bandwidth of the future ground-based 
Atacama Large Millimetre Array (ALMA; 
Ishiguro et al.\ 1994; 
Brown et al.\ 1996; Downes et al.\ 1996), although the current goal is for 
a 16-GHz bandwidth (Wootten 2000). The count predictions in 
the very wide  
atmospheric window between 150 to 250\,GHz is shown in Fig.\,3(a). 
This band  
cannot be observed simultaneously using heterodyne 
instruments, but could 
plausibly be covered using an advanced grating or Fabry--Perot spectrograph 
feeding a sensitive bolometer detector array. In Fig.\,3(b) the 
far-infrared/submillimetre band between 460 and 1500\,GHz is shown. This 
band is matched to the 
specified spectral range of the 
SPIRE Fourier Transform Spectrograph destined for the 
{\it FIRST} satellite (Griffin 1997; Griffin et al.\ 1998). 

\begin{figure*}
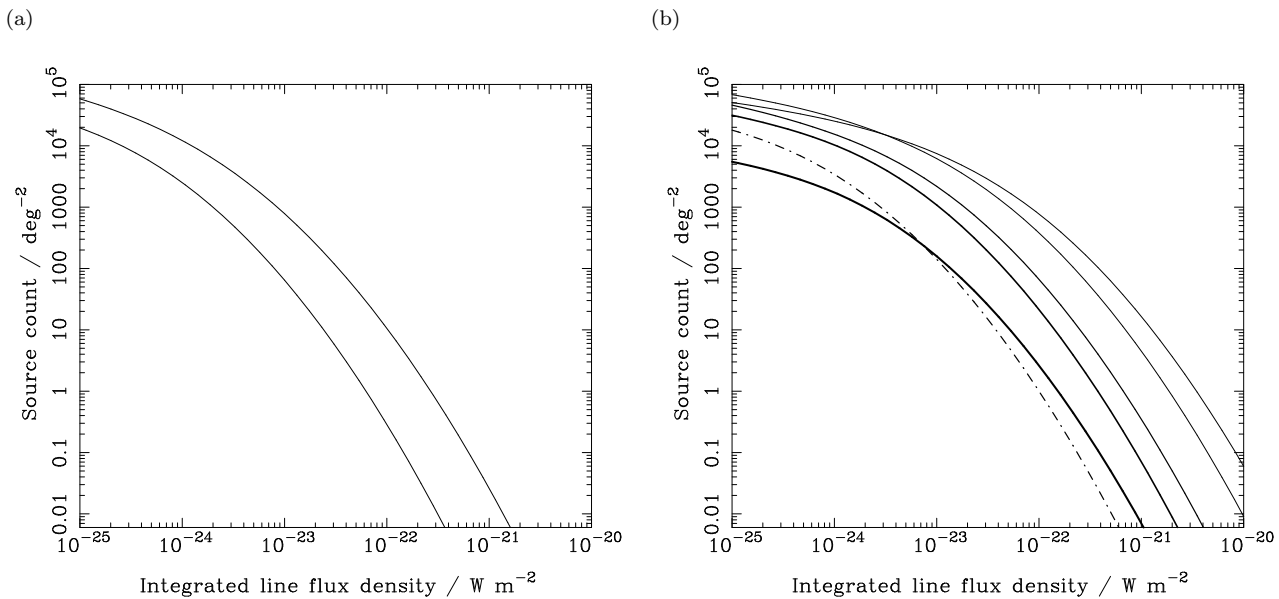
 
\begin{minipage}{170mm}
(a) \hskip 81mm (b)
\begin{center}
\epsfig{file=rev23.ps, width=7.05cm, angle=-90} \hskip 5mm
\epsfig{file=rev50.ps, width=7.05cm, angle=-90}
\end{center}
\caption{The line counts for galaxies at $z<10$ predicted in a 8-GHz band 
centred at (a) 23\,GHz and (b) 50\,GHz, corresponding to the K and Q 
bands. In (a) only the CO(1$\rightarrow$0) and 
CO(2$\rightarrow$1) lines are redshifted into the band. In (b) CO lines 
up to CO(5$\rightarrow$4) and the {[C\sc{i}]}$_{492\,{\rm GHz}}$ line are 
present. 
The line styles and thicknesses that 
correspond to the transitions are listed in Table\,1. 
$10^{-22}$\,W\,m$^{-2}$ is equivalent to 
0.13 and 0.06\,Jy\,km\,s$^{-1}$ at 23 and 50\,GHz respectively.
}
\end{minipage}
\end{figure*} 

\begin{figure*}
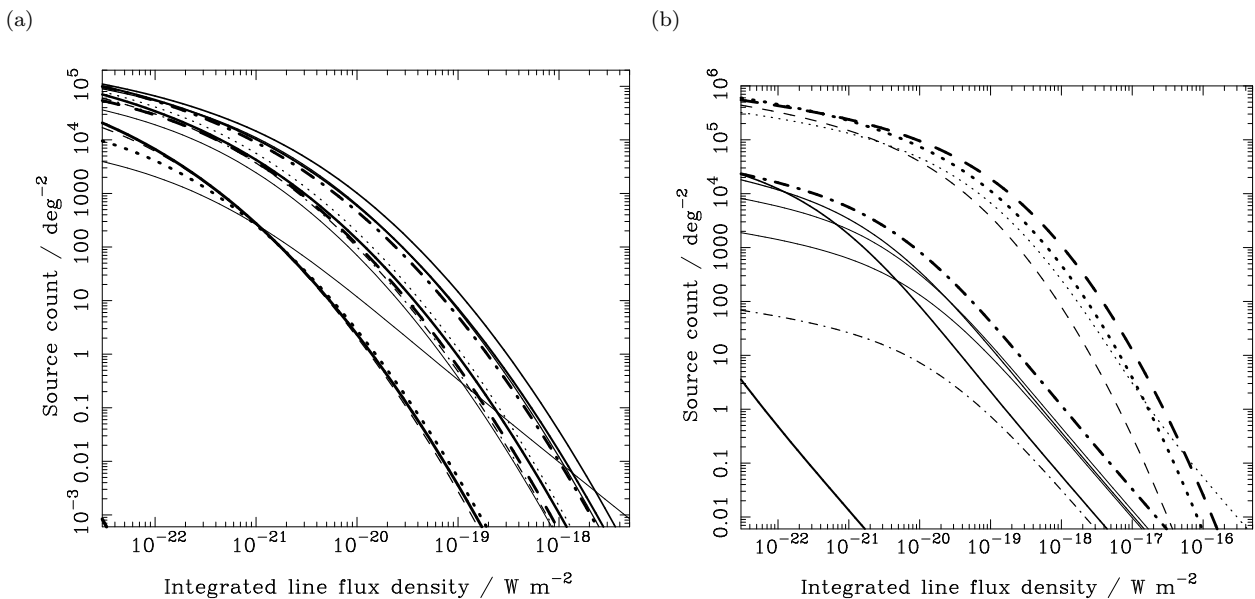

\begin{minipage}{170mm}
(a) \hskip 81mm (b)
\begin{center}
\epsfig{file=rev150.ps, width=7.05cm, angle=-90} \hskip 5mm
\epsfig{file=revSPIRE.ps, width=7.05cm, angle=-90}
\end{center}
\caption{The line counts predicted in very wide bands covering the spectral 
ranges between 150--250\,GHz (a) and 460--1500\,GHz (b). 
The band in (a) could be observed using a grating spectrograph feeding 
bolometer detectors. The 
band in (b) could be surveyed using the SPIRE-FTS 
instrument fitted to {\it FIRST}.  
The lowest 
CO transition shown in (a) is CO(2$\rightarrow$1). In (b) the lowest 
CO transition is CO(5$\rightarrow$4), and the counts are dominated by 
fine-structure lines. The line styles and thicknesses that correspond to 
each transition are listed in Table\,1. 
$10^{-20}$\,W\,m$^{-2}$ is equivalent to 1.5 and 
0.4\,Jy\,km\,s$^{-1}$ at 200 and 800\,GHz respectively. 
}
\end{minipage}
\end{figure*} 

\begin{figure*}
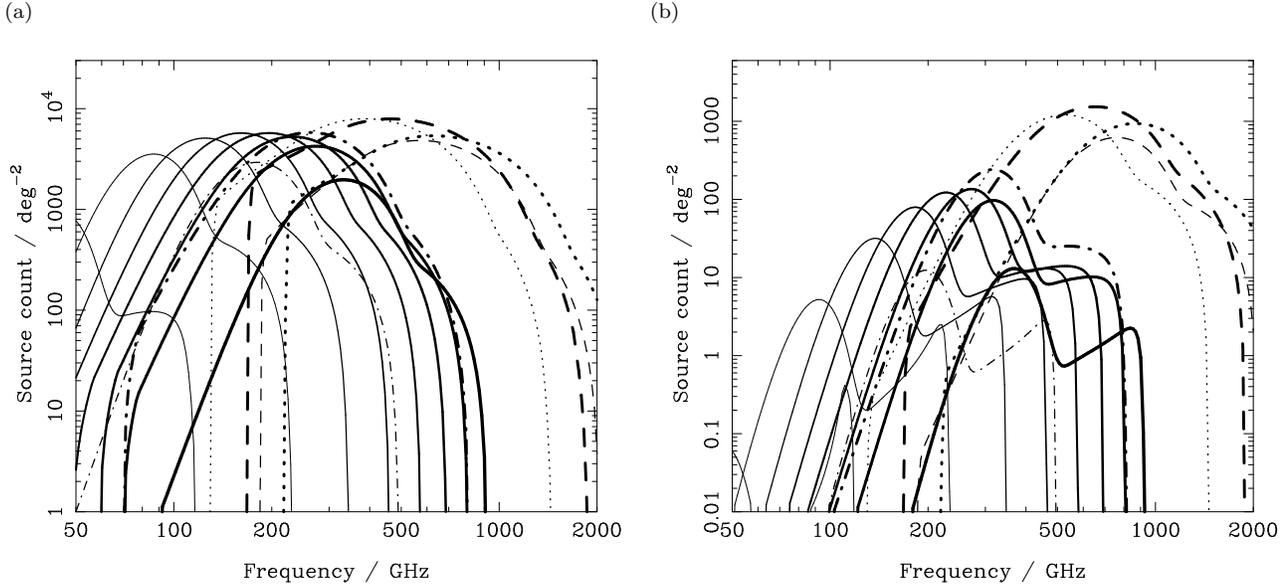

\begin{minipage}{170mm}
(a) \hskip 81mm (b)
\begin{center}
\epsfig{file=rev_freq-22.ps, width=6.95cm, angle=-90} \hskip 5mm
\epsfig{file=rev_freq-20.ps, width=6.95cm, angle=-90}
\end{center}
\caption{The surface density of line emitting galaxies brighter than (a) 
$10^{-22}$\,W\,m$^{-2}$ and (b) $10^{-20}$\,W\,m$^{-2}$, as a function 
of observing frequency, assuming an observing bandwidth of 8\,GHz. 
The line styles are listed in Table\,1. The 
sharp lower cutoff to the counts of fine-structure lines at low frequencies 
is 
attributable to the assumed upper limit to the redshift distribution of dusty 
galaxies, $z_0=10$. If gas-rich galaxies 
exist at $z > 10$, then they would be detectable at frequencies below these 
cutoffs. 
}
\end{minipage}
\end{figure*}

\subsection{The shape of the counts and optimal surveys}  

The integrated counts, $N(>S) \propto S^{-\alpha}$ presented 
in Figs 1 to 3 all have a characteristic form, with a relatively 
flat slope, $\alpha \simeq 0.3$, at faint flux densities, which  
rolls over to a steep decline, $\alpha \simeq 3$,  
at brighter fluxes. 
Note that the detection 
rate of galaxies in a survey is maximized at the  
depth at which $\alpha = 2$. Because the counts of 
lines presented here have a pronounced knee at a certain flux density, 
at which the 
value of $\alpha$ crosses $2$, this flux density is the optimal depth for a 
blank-field line survey, and surveys 
that are either shallower or deeper should be much less efficient. 

In Fig.\,4 the predicted counts of line-emitting galaxies are 
shown as a function 
of observing frequency between 50 and 2000\,GHz, at both faint and relatively 
bright values of the integrated line flux density $S$, $10^{-22}$ and 
$10^{-20}$\,W\,m$^{-2}$. 
The form of the curve for each emission line is determined by 
the form of evolution of far-infrared luminous galaxies specified in the 
modified Gaussian model. In any particular 
line, galaxies detected at higher frequencies are at lower redshifts. This effect 
accounts for the double-peaked distribution that is most apparent in the counts 
of bright galaxies shown in Fig.\,4(b). The broad peak at lower frequencies 
corresponds to the very luminous Gaussian burst of activity at $z \simeq 1.7$, 
while less luminous low-redshift galaxies account for the 
second sharper peak at higher frequencies. The double peak is more noticeable 
for the brighter counts, to which  
the contribution of low-redshift galaxies is more 
significant.

Based on the predicted counts of 
CO lines, which are probably quite reliable, 
the most
promising frequency range in which to aim for CO line detections is
200--300\,GHz, regardless of whether faint or bright CO lines are being sought.
At frequencies greater than about 500\,GHz 
the counts are expected to be 
dominated by fine-structure lines, and although these counts are still 
very uncertain, the 
detectability of   
fine-structure line emitting galaxies is expected to 
be maximized at a frequency 
between about 400 and 800\,GHz. Future blank-field 
searches for fine-structure 
emission from distant galaxies should probably be concentrated in 
this range.

\begin{figure*}
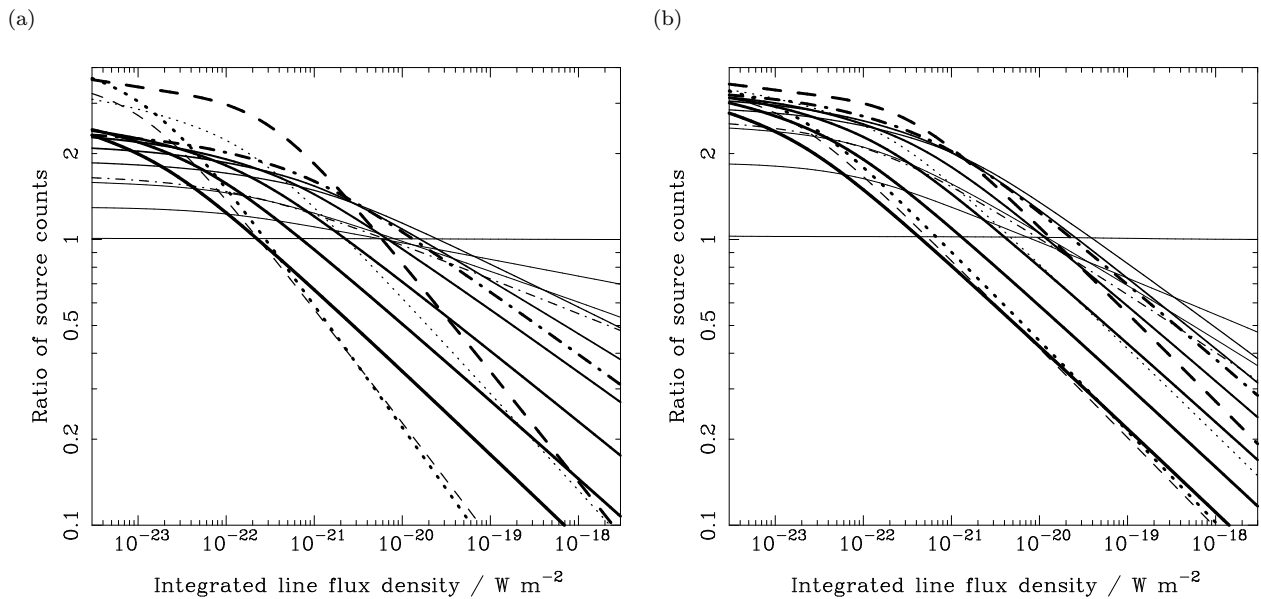

\begin{minipage}{170mm}
(a) \hskip 81mm (b)
\begin{center}
\epsfig{file=gra_ratio_53_230_Open.ps, width=7.05cm, angle=-90} \hskip 5mm
\epsfig{file=gra_ratio_53_230_Lambda.ps, width=7.05cm, angle=-90}
\end{center}
\caption{The effects of different world models on the predictions of CO and 
fine-structure line counts, for demonstration in a narrow band centred on 
230\,GHz (see Fig.\,2). The line styles and thicknesses that 
correspond to each transition are listed in Table\,1. 
The results are presented as ratios between the counts 
in two world models with $\Omega_0 < 1$, and the predictions of an 
Einstein--de Sitter model. In (a) the cosmological parameters are 
$\Omega_0=0.3$ and $\Omega_\Lambda=0.0$; in (b) $\Omega_0=0.3$ and 
$\Omega_\Lambda=0.7$. The differences due to the world model are by factors 
of a few, at worst comparable with the level of uncertainty in the count 
predictions that is attributable to the uncertain excitation conditions in the ISM. 
}
\end{minipage}
\end{figure*} 

\subsection{The effects of CO excitation} 

The three different sets of values of $f_{\rm line}$ listed for CO 
transitions in 
Table\,1 were derived assuming different excitation conditions in the ISM. As 
the detailed astrophysics of gas-rich high-redshift galaxies is very uncertain 
and only weakly constrained by observations, there is an inevitable uncertainty in 
the counts predicted in Figs 1--4. 

The most significant differences between the values of $f_{\rm line}$ 
predicted in the LVG, 38-K and 53-K 
models occur in the higher $J$ lines, for which the effect of the 
higher excitation temperature and the subthermal excitation of lines in 
the LVG model is greater. The increase is most significant 
at bright flux densities. If the alternative values of $f_{\rm line}$ listed 
in the thermal equilibrium models are used to predict counts similar to 
those in Figs\,1--4, then for $J \le 7$, at 
interesting line flux densities of about $10^{-22}$\,W\,m$^{-2}$,
the counts differ by less than a factor of about 3. This  
uncertainty is greater than that in the surface density of high-redshift 
submillimetre-selected galaxies that were used to normalize the 
model of galaxy evolution, and so the counts predicted in Figs\,1--4
should 
be reliable to within a factor of about 3. Note that the counts in 
Fig.\,4 indicate that the CO(5$\rightarrow$4) line is likely to be provide 
the most significant contribution to the counts, and so the degree of 
excitation of lines with $J \ge 7$ 
is not likely to affect the detection rate in future
blank-field line surveys. 

\subsection{The effects of cosmology} 

In Fig.\,5 the 230-GHz counts expected in three different world models 
are compared, based on the same   
underlying population of distant dusty galaxies. In world models  
with $\Omega_0=0.3$ and either $\Omega_\Lambda = 0.0$ or 0.7 
the faint counts are expected 
to be greater, and the bright counts are expected to be less, as compared with 
the predictions of an Einstein--de Sitter model. 
The count changes as a
function of cosmology by a factor of about 10 in the most abundant lines 
over
the wide range of flux 
densities presented in Fig.\,5, which correspond to a range of count values 
between about 10$^4$ and 10$^{-3}$\,deg$^{-2}$. 
At a 
flux density of $10^{-20}$\,W\,m$^{-2}$, which is likely to be similar to the 
limiting depth of future submillimetre-wave line surveys, the counts are 
modified by only a factor of about 2 by changing the world model. Hence, at 
present the uncertainties in the excitation conditions in the ISM  
are expected to be greater than those introduced by 
uncertainties in the cosmological 
parameters. 

\section{Line observations}

\subsection{Millimetre-wave 
interferometer arrays} 

Four millimetre-wave interferometer 
arrays are currently operating: the BIMA 
array with ten 6-m antennas; the IRAM array 
with five 15-m antennas; the Nobeyama Millimeter Array
with six 10-m antennas; and the OVRO Millimeter Array with six 
10.4-m antennas. These instruments operate at wavelengths 
longer than 1\,mm, and their fields of view and sensitivities 
translate into similar mapping speeds to equivalent flux density 
levels. 
For example, in a 20-hr integration at 
a frequency of 90\,GHz, the OVRO Millimeter 
Array is able to reach a $5\sigma$
sensitivity limit of about $5 \times 10^{-21}$\,W\,m$^{-2}$ across a 
1-GHz band in a 1-arcmin$^2$ primary beam. 

The counts of lines expected in a 90-GHz observation with a 1-GHz 
bandwidth are shown in Fig.\,1(a) as a function of integrated line flux 
density. At this limiting flux density, a count of 
about 2\,deg$^{-2}$ is expected, and so the 
serendipitous detection of an emission line in a blank-field survey would be 
expected to occur after 
approximately $4 \times 10^4$\,hr. In 230-GHz observations 
with the same 1-GHz bandwidth, which are possible in good 
weather, the sensitivity required for a 5$\sigma$ detection in a 20-h 
integration 
is about $2 \times 10^{-20}$\,W\,m$^{-2}$, 
in a 0.25-arcmin$^2$ primary beam. The 230-GHz 
count expected at this depth, after correcting the results shown in 
Fig.\,1(b) for the narrower bandwidth, is about 50\,deg$^{-2}$. A 
serendipitous line detection would thus be expected about 
every 6000\,hr. 
Thus, while known high-redshift galaxies can 
certainly be reliably detected in reasonable integration times using these 
arrays, blank-field surveys for emission lines 
are not currently practical. 

The detection of redshifted CO lines at a 90-GHz flux density of about 
$10^{-20}$\,W\,m$^{-2}$ from two submillimetre-wave continuum 
sources, with a surface density of several 100\,deg$^{-2}$ (Blain 
et al.\, 1999b, 2000), using the OVRO array 
(Frayer et al.\ 1998, 1999) is consistent with the 90-GHz line count 
of about 2\,deg$^{-2}$ predicted in Fig.\,1(a) at this flux density.
This is because observations in 
many tens of 1-GHz bands 
would have been required to search for a CO emission line from 
these galaxies without an optical redshift, and so 
the source count of lines at the detected flux density is expected to
be about two orders of magnitude less than that of the dust 
continuum-selected galaxies. 

The large ground-based millimetre/submillimetre-wave 
interferometer array ALMA  
will provide excellent subarcsecond angular resolution and a large collecting 
area for observations of submillimetre-wave continuum and line radiation. 
Based on the performance described for ALMA at 230\,GHz by Wootten (2000), 
a 300-km-s$^{-1}$ line with an integrated 
flux density of $5 \times 10^{-22}$\,W\,m$^{-2}$ could be detected at $5\sigma$
significance, but not resolved, anywhere within a 16-GHz band in about 1\,hr 
in the 0.15-arcmin$^2$ primary beam.  
The surface density of lines brighter than this flux density is expected to 
be about $2.5 \times 10^4$\,deg$^{-2}$ (Fig.\,1b; see also Blain 2000), and so 
a detection rate of about 1.7\,hr$^{-1}$ would be expected. The knee in the 
counts in Fig.\,1(b), indicating the most efficient survey depth, is at a flux 
density greater than about $10^{-20}$\,W\,m$^{-2}$, a depth reached in an 
integration time of about 10\,s at 5$\sigma$ significance. Hence, making a 
large mosaic map at this depth, covering an area of 0.017\,deg$^2$\,hr$^{-1}$ 
should maximize the detection rate, which would then be about 15 galaxies per 
hour, neglecting scanning overheads incurred in the mosaiking process. This 
detection rate would allow large samples of line-emitting high-redshift 
galaxies to be compiled rapidly using ALMA. The performance of ALMA in 
line searches is discussed in more detail by Blain (2000).

\subsection{Ground-based single-antenna telescopes} 

At present the heterodyne spectrographs fitted to the  
JCMT, the Caltech Submillimeter Observatory (CSO), and the 
IRAM 30-m antenna are not sufficiently sensitive to allow a blank-field 
survey to search for distant line emitting galaxies. For example, the 
230-GHz receiver at the JCMT, which is the least 
susceptible to atmospheric noise, can reach a 5$\sigma$ sensitivity of about 
$2 \times 10^{-19}$\,W\,m$^{-2}$ in a 1-hr observation in a 1.8-GHz band 
centred on 230\,GHz. At this flux density the surface density of CO line 
emitting galaxies is expected to be $\ls 10$\,deg$^{-2}$ (see Fig.\,1b), 
and so because 
the beam area is $2.4 \times 10^{-5}$\,deg$^2$, many tens of 
thousands of hours of observation would be required to detect a source 
serendipitously. 
The development of wide-band 
correlators, such as the 3.25-GHz WASP (Isaak, Harris \& Zmuidzinas 1999), 
will improve the performance of single-antenna telescopes significantly for the 
detection of faint lines in galaxies with a known redshift, 
but unless bandwidths 
are increased by at least an order of magnitude, blank-field line searches from 
the ground will not be practical. 

\subsubsection{Future instrumentation} 

At present, submillimetre-wave emission lines from a particular distant 
galaxy can only be detected if an accurate redshift has been determined, 
because the instantaneous bandwidth of receivers is narrow, and
accurate tuning is required for a target line to
be observed within the available band.
For galaxies and AGN detected in blank-field submillimetre continuum
surveys, obtaining a spectroscopic redshift requires a 
very considerable investment of observing time
(see, e.g., Smail et al.\ 1999a). However, 
if a much 
wider 
bandwidth of order 
$115\,{\rm GHz} / (1+z)$ could be observed simultaneously,
then a CO line would always lie within the observing band from a galaxy at 
redshift $z$. 

There are currently two ways to increase the instantaneous 
bandwidth of a ground-based millimetre-wave telescope, both based on 
bolometer detectors rather than heterodyne mixers. Either a resonant cavity, 
such as a Fabry--Perot, or a diffraction grating could be used to feed an array 
of bolometer detectors. The potential of such instruments are discussed in 
the subsections below. 

\subsubsection{A Fabry--Perot device: SPIFI} 

SPIFI is a $5\times5$ element Fabry-Perot interferometer, 
for use at frequencies from 460 to 1500\,GHz on both the 
1.7-m AST-RO 
telescope at the South Pole and the 15-m JCMT (Stacey et al.\ 1996; 
Bradford et al.\ 1999, in preparation). On 
AST-RO, in the 350-$\mu$m (850-GHz) atmospheric window, the field of view 
of the instrument is about 25\,arcmin$^2$. At a coarse 300-km-s$^{-1}$ 
resolution, a bandwidth of about 8\,GHz can be observed 
to a 
$5\sigma$ detection threshold of about $3.2 \times 10^{-17}$\,W\,m$^{-2}$ in a 
1-hr  
integration. On the 15-m JCMT, at the same frequency and for the 
same bandwidth, the field of view and the 
$5\sigma$ detection threshold are both less, 
about 0.3\,arcmin$^2$ and $8 \times 10^{-19}$\,W\,m$^{-2}$ 
respectively. 

From Fig.\,1(d), 
the count at flux densities brighter than the 5$\sigma$ detection threshold 
for the JCMT are expected to be about 30\,deg$^{-2}$, and to be dominated by 
[C{\sc ii}]-emitting galaxies, for which the counts are very 
uncertain. Hence 
a serendipitous detection would be expected about 
every 400\,hr, and so  
blank-field line surveys using SPIFI are on the
threshold of being practical. 
A second-generation instrument with a wider field of view
should be capable of making 
many detections in a reasonable integration time.

\subsubsection{A millimetre-wave grating spectrograph}

An alternative approach to obtaining a wide simultaneous bandwidth 
would be to build a grating spectrograph to disperse the signal from 
a 10-m ground-based telescope 
onto linear arrays of bolometers. 
Such an instrument would be 
able to observe a reasonably large fraction of the clear 150--250\,GHz 
atmospheric window simultaneously, at a 
background-limited $1\sigma$ sensitivity of 
about $10^{-20}$\,W\,m$^{-2}$ in a 1-hr 
observation, which is uniform to 
within a factor of about two across this spectral range. 
The predicted counts 
of lines in this spectral range are of 
order 100\,deg$^{-2}$ 
at this depth (Fig.\,3a), and are 
expected to be dominated by CO lines, for which the 
predicted counts should be reasonably accurate. At 200\,GHz the field of view of 
a 10-m telescope is about 0.2\,arcmin$^2$, and so a 
5$\sigma$ detection rate of about 
0.05\,hr$^{-1}$ would be expected in a blank-field line survey. A grating 
instrument with these specifications is thus on the 
threshold of being useful for conducting such a survey.

It would also be very valuable for detecting 
CO lines emitted by the high-redshift galaxies detected in 
dust continuum surveys whose positions are 
known to within about 10\,arcsec. An instantaneous bandwidth of order 100\,GHz 
would accommodate either 2 or 3 CO lines for a $z>2$ galaxy, and so a 
spectroscopic redshift could be determined without recourse to radio, optical 
or near-infrared telescopes. For example, the $z=2.8$ 
SCUBA galaxy SMM\,J02399$-$0136 
has an integrated flux density of $1.5 \times 10^{-21}$\,W\,m$^{-2}$ in 
the CO(3$\rightarrow$2) transition (Frayer et al.\ 1998), 
and all the CO($J$+1$\rightarrow$$J$) transitions 
with $J$=5, 6, 7 and 8 fall within the 150--250\,GHz spectral range. Based on 
the LVG model (Table\,1), these 
lines are expected to have integrated flux densities of $3.0$, $3.6$, $3.2$ 
and $1.0 \times 10^{-20}$\,W\,m$^{-2}$ respectively. Thus, in a 5-hr integration in 
the field of SMM\,J02399$-$0136, all but the CO(8$\rightarrow$7) line could be 
detected using a background-limited grating spectrograph, 
allowing its redshift to 
be determined unequivocally in a comparable time to that 
required to obtain an optical redshift using a 4-m 
telescope (Ivison et al.\ 1998b). There are two further advantages. First, the 
CO-line redshift would be the redshift of 
cool gas in the ISM of the galaxy, and not that of optical 
emission lines, which are typically blueshifted by several 
100\,km\,s$^{-1}$ with 
respect to the ISM. Secondly, the ratios of the luminosities of the detected 
CO lines would reveal information about the physical conditions in the cool 
ISM where star-formation takes place. 

\subsection{Ground-based centimetre-wave telescopes} 

Low-$J$ CO lines are redshifted into the centimetre waveband for redshifts 
$z < 10$. Instruments operating in this waveband include the VLA and 
the Green Bank Telescope (GBT).

Currently, the VLA can observe spectral lines in a very narrow 87.5-MHz 
band, centred in the K band (22--24\,GHz) and Q band  
(40--50\,GHz). In the K band, Ivison et al.\ (1996) 
attempted to detect CO(1$\rightarrow$0) emission from
the environment of a dusty radio galaxy at $z=3.8$, and in 12\,hours they 
obtained a 
$5\sigma$ upper limit of about 
$4 \times 10^{-23}$\,W\,m$^{-2}$ in a 2-arcmin beam. Because of the 
very narrow bandwidth, the 
detection rate of CO line-emitting galaxies with the K-band VLA in a 
blank-field survey is expected to be about 1\,yr$^{-1}$.
However, by 2002 fibre-optic links and a new correlator 
will be installed, increasing the simultaneous bandwidth greatly to 
8\,GHz, and improving the $5\sigma$ sensitivities to $1.3 \times 10^{-23}$ 
and $8.0 \times 10^{-23}$\,W\,m$^{-2}$ in 12-hr K- and Q-band 
integrations respectively. 
The counts of line-emitting galaxies at $z\le10$ in these bands are
shown in Fig.\,2. At these sensitivities, 
K- and Q-band counts of about 500 and 2000\,deg$^{-2}$ are expected  
respectively, each corresponding to a detection rate of about 0.03\,hr$^{-1}$. 
The knee in the counts at which the detection rate is maximized is expected 
at rather similar flux densities of about 
$3 \times 10^{-23}$ and $2 \times 10^{-22}$\,W\,m$^{-2}$ in the K and 
Q bands respectively, depths which can be reached in about 140- and 
190-min integrations and at which the counts are expected to be about 
100 and 400\,deg$^{-2}$ 
respectively. The most efficient detection rates in the K and Q bands 
are thus expected to be about 0.04 and 0.03\,hr$^{-1}$. 
Hence, low-$J$ blank-field CO-line surveys will 
be practical using the upgraded VLA. High-redshift galaxies with luminosities 
comparable to the SCUBA galaxies detected by Frayer et al.\ (1998, 1999) 
should be readily detectable, as their integrated flux densities in the 
CO(1$\rightarrow$0) line would be of order $7 \times 10^{-23}$ and 
$4 \times 10^{-22}$\,W\,m$^{-2}$ when redshifted into the K and Q bands, 
from $z=4$ and 1.3 respectively.

The 100-m clear-aperture GBT will operate with a 
3.2-GHz bandwidth in the K and Q bands, reaching 5$\sigma$ sensitivity 
limits in 1-hr integrations of about 1.3 and $5 \times 10^{-22}$\,W\,m$^{-2}$ 
respectively. These sensitivities make the GBT even more suitable for 
detecting low-$J$ CO lines 
from known high-redshift galaxies than the VLA, 
but the subarcminute field of view of the 
GBT is too small to allow a practical blank-field survey. At these 
sensitivity limits,  
only $2 \times 10^{-5}$ and $6 \times 10^{-4}$ 
lines per beam are expected in the K and Q bands respectively. 

\subsection{Air- and space-borne 
instruments} 

The 3.5-m space-borne telescope {\it FIRST} (Pilbratt 1997) will carry the 
HIFI far-infrared/submillimetre-wave spectrometer (Whyborn 1997) and the 
SPIRE bolometer array camera (Griffin et al.\ 1998). Both of these 
instruments operate at frequencies for which fine-structure lines are expected 
to dominate the counts of line-emitting galaxies, and so the number of 
detectable galaxies is necessarily uncertain. 

The spectral coverage of HIFI extends from 500 to 1100\,GHz. The 
instrument will have a 4-GHz 
bandwidth, a 0.4-arcmin$^2$ field of view and a 5$\sigma$ sensitivity of 
about $5 \times 10^{-19}$\,W\,m$^{-2}$ in a 1-hr integration at 650\,GHz. 
The SPIRE Fourier Transform Spectrograph (SPIRE-FTS) 
will provide a spectroscopic 
view of a 2-arcmin-square field at all frequencies between 460\,GHz and 
1.5\,THz simultaneously.  
The 5$\sigma$ detection threshold of a line in a 1-hr 
integration is expected to 
be about $1.5 \times 10^{-16}$\,W\,m$^{-2}$. 
The slope of the counts shown in Figs\,1(d) and 3(b), which are 
based on the estimated, and very uncertain, counts of [C{\sc ii}] lines, 
indicate that the 
most efficient survey depth 
is greater than 
$10^{-18}$\,W\,m$^{-2}$ for HIFI and about  
$5 \times 10^{-18}$\,W\,m$^{-2}$ for SPIRE-FTS. At a depth of 
$5 \times 10^{-18}$\,W\,m$^{-2}$,   
the optimal detection rates using HIFI and SPIRE-FTS are expected to be
about $6 \times 10^{-3}$ and $7 \times 10^{-5}$\,hr$^{-1}$ respectively, 
and so
the detection 
rate of line-emitting galaxies using FIRST is expected to be quite low. 

The currently quoted sensitivities of 
{\it FIRST}-HIFI are about seven times better than those of the heterodyne 
instruments attached to the 2.5-m SOFIA airborne telescope 
(Becklin 1997; Davidson 2000). Thus, when the 
larger field of view of SOFIA is taken into account, 
a line-emitting galaxy could 
be detected using SOFIA about every 4500\,hr. Hence, SOFIA could not carry 
out a successful blank-field line emission survey. Note, however, that the 
instruments aboard SOFIA will be upgraded throughout its 20-year 
lifetime, and so the development of 
multi-element detectors and innovative wide-band 
spectrographs
may change this situation.

The proposed space-borne far-infrared  
interferometer {\it SPECS} (Mather et al.\ 1998) has very 
great potential for resolving and obtaining very high signal-to-noise 
spectra of high-redshift galaxies. Although predictions of the detectability of 
fine-structure lines at frequencies greater than 600\,GHz are very 
uncertain, with a field of view of about 0.25\,deg$^{-2}$ at 650\,GHz 
and a $5\sigma$ sensitivity of about 10$^{-18}$\,W\,m$^{-2}$ in a 
24-hr integration, about 5 [C{\sc ii}] galaxies could be detected per day 
using {\it SPECS} in a 8-GHz-wide band (Fig.\,1d). Hence, 
although {\it SPECS} is predominantly an instrument to study known galaxies 
in great detail, its sensitivity is sufficient to carry out successful 
blank-field line surveys.

\subsection{Summary of prospects for line surveys} 

The most promising instruments for future surveys of CO and 
fine-structure lines from high-redshift galaxies are the 
ALMA interferometer array in the 140- and 230-GHz bands, which should be 
able to detect about 15 galaxies an hour in a survey to a 
line flux density of 
$10^{-20}$\,W\,m$^{-2}$ in a 16-GHz-wide band. 
Other instruments for which detection rates are expected to 
exceed one source per hundred hours are the future 
{\it SPECS} space-borne interferometer, which is likely to 
detect of order of five fine structure line-emitting galaxies per 
day, and a ground-based, wide-band, background-limited millimetre-wave 
grating spectrograph, which should 
be able to detect a line-emitting galaxy every 20\,hr, the same rate as 
the upgraded K- and Q-band VLA. 

\section{Conclusions} 

We draw the following conclusions 

\begin{enumerate} 

\item We have predicted the surface densities of high-redshift galaxies that 
emit molecular rotation and atomic fine-structure line radiation that is
redshifted into the
millimetre/submillimetre waveband. The results depend on both the excitation 
conditions in the ISM of the galaxies and the evolution of the 
properties of distant gas-rich 
galaxies. We incorporate the latest results of CO-line observations and the 
redshift distribution of galaxies discovered in submillimetre-wave continuum 
surveys to provide a sound observational basis for these predictions.

\item The predicted counts of CO line-emitting galaxies are probably accurate 
to within a factor of about 5 for lines with $J \ls 7$. The uncertainties are
inevitable, and caused by the lack of knowledge of both the cosmological model 
and the excitation state of the emitting gas in the ISM of high-redshift
galaxies. 

\item The predictions of the counts of atomic fine-structure lines in
high-redshift ultraluminous galaxies are more weakly constrained by
observations, and we present likely order-of-magnitude estimates of the 
counts in six such lines. 

\item The most efficient frequency for a survey that aims to detect CO 
emission is probably in the range 200--400\,GHz, which includes the 
230- and 350-GHz atmospheric windows. The optimal frequencies for a 
fine-structure line survey are probably in the range 400--800\,GHz, 
which includes the 670- and 850-GHz atmospheric windows.

\item There are excellent prospects for using a range of millimetre- and
submillimetre-wave instruments that are currently under development 
to conduct blank-field surveys for the redshifted emission lines from the ISM 
in high-redshift galaxies. ALMA will probably prove the most capable facility 
for blank-field surveys, and should detect up to about 15 line-emitting 
galaxies per hour. The most efficient survey is likely to be made in the 
140- and 230-GHz bands to a  
an integrated line flux density of a few $10^{-20}$\,W\,m$^{-2}$. 
The required sensitivities could also be achieved by 
future wide-band spectrometers on single-antenna telescopes. Low-$J$ 
CO transitions will be detectable using the upgraded VLA  
at longer millimetre and centimetre wavelengths.

\item A wide range of millimetre/submillimetre-wave lines should be detected in 
the spectra of galaxies selected in other wavebands, and in blank-field dust 
continuum surveys. Redshifts for millimetre/submillimetre-selected galaxies 
could be determined directly in these wavebands if a wide bandwidth of 
order $100 / (1+z)$\,GHz was available, with a centre frequency of about 
200\,GHz.

\end{enumerate}

\section*{Acknowledgements} 

The results in this paper are based on the properties of 
the SCUBA lens survey galaxies detected at the Owens Valley Millimeter Array 
in collaboration with Aaron Evans and Min Yun. The core of the 
SCUBA lens survey was carried out by Ian Smail, Rob Ivison, AWB and 
Jean-Paul Kneib. We thank the referee Paul van der Werf for his careful 
reading of the manuscript and valuable comments, and also 
Jackie Davidson, Kate Isaak, Rob Ivison, Richard Hills, Brett Kornfeld, 
Malcolm Longair, Phil Lubin, Kate Quirk, John Richer and Gordon Stacey 
for helpful conversations and comments. AWB -- the Raymond \& Beverly 
Sackler Foundation Research Fellow at the IoA, Cambridge -- 
gratefully acknowledges generous support from the Raymond \& Beverly
Sackler Foundation as part of the Deep Sky Initiative programme at the IoA. 
AWB thanks the Caltech AY visitors program for support while this work 
was conducted.

\end{document}